\documentclass[aps,prl,superscriptaddress,twocolumn,showpacs,amsmath,amssymb]{revtex4}

\usepackage{color}
\usepackage{graphicx}
\usepackage{bm}

\begin{document}

\title{Odd momentum pairing and superconductivity in vertical graphene heterostructures}
\author{Francisco Guinea}
\affiliation{Instituto de Ciencia de Materiales de Madrid. Consejo Superior de Investigaciones Cient{\'\i}ficas. Sor Juana In\'es de la Cruz 3. E-28049 Madrid. Spain.}
\author{Bruno Uchoa}
\affiliation{Department of Physics and Astronomy, University of Oklahoma, Norman, OK 73069, USA}
\begin{abstract}
Vertical graphene heterostructures made up of graphene layers separated by BN spacers allow for novel ways of tuning the interactions between electrons. We study the possibility of electron pairing mediated by modified repulsive interactions. Long range intravalley and short range intervalley interactions give rise to different anisotropic phases. 
We show that a superconducting state with gaps of opposite signs in different valleys, an odd momentum pairing state, can exist above carrier densities of $5-10 \times 10^{13} {\rm cm}^{-2}$. The dependence of the transition temperature on the different parameters of the devices is studied in detail.
\end{abstract}
\maketitle
\section{Introduction}
Graphene has a number of electronic properties which can be tuned\cite{Netal04,Netal05}. Changes in the carrier concentration, nature of the substrate, and other properties of the environment can modify the electron-electron interaction\cite{NGPNG09,KUPGN12}.
Graphene structures, made by combining graphene layers separated by hexagonal boron nitride\cite{Metal11,Petal11,Letal11,Betal11,Betal12,Getal12}  show a number of interesting features, related to the interactions between electrons in different layers. These devices are highly tunable, as the carrier concentration in the graphene layers, and the distance between them can be varied independently.

\begin{figure}
\begin{center}
\includegraphics[width=0.6\columnwidth]{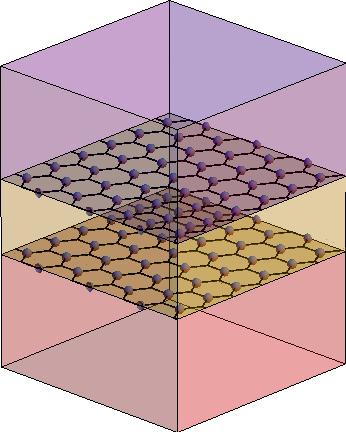}
\caption[fig]{\label{sketch}Sketch of the heterostructure considered in the text. Two graphene mono- or bilayers are embedded into three dielectric media. The three dielectrics need not have the same dielectric constant. The device is characterized by the carrier densities in the graphene layers, $\rho_t$ and $\rho_b$, the distance between layers, $d$, and the three dielectric constants, $\epsilon_1 = \epsilon_t , \epsilon_2 = \epsilon_m$ and $\epsilon_3 = \epsilon_b$.}
\end{center}
\end{figure}

We study here the possibility of superconductivity in these devices, arising from a suitably modulated electron-electron interaction. Superconductivity in metallic multilayers due to electron-electron interactions was proposed long ago\cite{F68}, as these systems exhibit an acoustic plasmon, which can mediate an attractive interaction  between the electrons. The model has been generalized to graphene near a metal surface\cite{UN07}. Alternatively, a repulsive electron-electron interaction, if its spatial dependence is suitably modulated, can lead to superonductivity\cite{KL65}. This possibility has been extensively studied in many materials (see, for instance\cite{R87,BD89,MD89,K91,GGV97,K02,BMK03,EB09,RK11,AK11}). These studies include carbon nanotubes\cite{PG06}, graphene and graphite at certain dopings or stacking arrangements\cite{BD07,G08,EB09,LOS10,LS10,NLC11,KHV11,EE11,VTGM12,ZQS12}, and combinations of two dimensional electron gases\cite{BD89,BMK03,AK11} (2DEG).

Experimentally, a number of graphene based compounds exhibit superconductivity. The graphite intercalation compounds\cite{Hetal65,DD81,DD02,Wetal05,Eetal05} show superconductivity up to temperatures $T_c \approx 11$K, and carrier concentrations in the graphene layers of $n \sim 1-4 \times 10^{14} {\rm cm}^{-2}$. A number of doped fullerenes have a superconducting phase\cite{H91,Getal08} with the highest critical temperature $T_c \sim 38$K. The carrier concentration at which superconductivity is found, normalized to the number of carbon atoms, is similar to the concentrations in the graphite intercalation compounds. The origin of superconductivity in doped fullerenes at high pressures is not completely understood, and it is likely that electron-electron interactions play a role\cite{CFCT09}.

The carrier density and interlayer distance in graphene heterostructures determine the features of the electron-electron interaction, which can be changed over a wide range. This opens the possibility of superconducting phases at not too large carrier densities and without the requirement of special features in the density of states. The analysis of superconducting instabilities is the main goal of this work.  The next section discusses in detail the effective interaction. Then, we analyze possible superconducting instabilities using the Kohn-Luttinger framework, and we also characterize the most likely superconducting phase.  We find that an odd momentum  paring state, with gaps of opposite signs in the different cones, is favored by intervalley scattering at large densities and could be the leading instability.  We present next the dependence of the transition temperature on the main parameters of the model, the nature of the superconducting phase, and also the relation to the formation of an excitonic condensate in an artificial bilayer with electrons and holes.

\section{Effective interaction in graphene heterostructures}

We consider two graphene layers, each of which can be single layer graphene or bilayer graphene with carrier concentration $n_t$ and $n_b$, where $t$ and $b$ denote the top and bottom layers. They are separated at a distance $d$. The top, middle, and bottom dielectrics have static dielectric constants $\epsilon_1 , \epsilon_2$ and $\epsilon_3$, as shown in Fig.~\ref{sketch}. The parameters used are: nearest neighbor hopping  $\gamma_0 \approx 2.7$ eV, distance between nearest neighbor carbon atoms $a \approx 1.4$ \AA, Fermi velocity $\hbar v_F = (3 \gamma_0 a )/2$, fine structure constant of graphene $\alpha = e^2 / ( \hbar v_F ) \approx 2.5$, and hopping between nearest neighbors in different layers $\gamma_1 \approx0.4$ eV.

\begin{figure}
\begin{center}
\includegraphics[width=0.6\columnwidth]{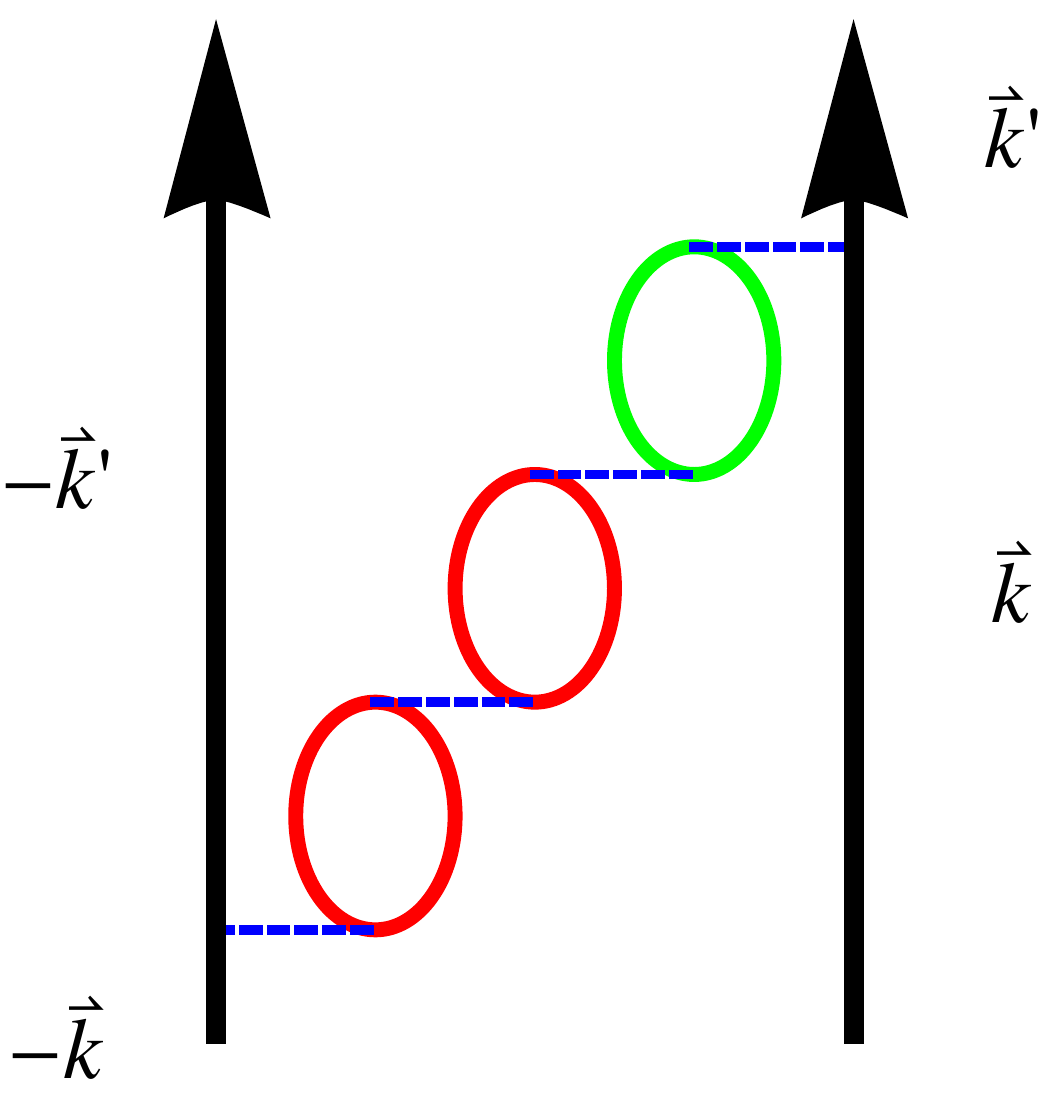}
\caption[fig]{\label{KL_propagator}Diagram included in the calculation of the effective interaction, $V ( q )$. Red and blue bubbles stand for the polarizabilities in the two layers in the heterostructure. The dashed lines describe the Coulomb interaction, which can be intra- or interlayer.}
\end{center}
\end{figure}

We assume that Cooper pairs are formed by electrons which occupy states related by time inversion symmetry\cite{symmetry}. We consider processes in which the Cooper pairs are scattered within one valley, and between valleys, as shown in Fig.~\ref{valleys}. We study first the effect of the scattering of Cooper pairs within one valley, shown in Fig.~\ref{valleys} a).
We approximate the long wavelength electron-electron interaction between the layers by its RPA value, see Fig.~\ref{KL_propagator}. This approximation is justified when the number of fermion species, $N_f$, is large. In our case, $N_f=4$.

The effective interaction for small momenta, $q a \rightarrow 0$, in the top layer is given by
\begin{widetext}
\begin{align}
v_{tt} ( q ) &= v_c ( q ) \epsilon_{tt}^{-1} \frac{1 - v_c ( q ) \chi_b ( q ) \left( \epsilon_{bb}^{-1} ( q ) - \epsilon_{tb}^{-1} ( q ) e^{- q d} \right)}
{1  - v_c ( q ) \left[  \epsilon_{tt}^{-1} \chi_t ( q ) + \epsilon_{bb}^{-1} \chi_b ( q )\right] + v_c^2 ( q ) \chi_t ( q ) \chi_b ( q ) \left[ \epsilon_{tt}^{-1} ( q ) \epsilon_{bb}^{-1} ( q ) - \epsilon_{tb}^{-2} e^{-2 q d} \right]}
\label{veff}
\end{align}
\end{widetext}
where $v_c ( q ) = 2 \pi e^2 / q$ is the Coulomb potential, and
\begin{align}
\epsilon_{tt} ( q ) &= \frac{e^{2 q d} \left( \epsilon_3 + \epsilon_2 \right) \left( \epsilon_1 + \epsilon_2 \right) - \left( \epsilon_3 - \epsilon_2 \right) \left( \epsilon_1 - \epsilon_2 \right)}{2 \left[ \left( 1 + e^{2 q d} \right) \epsilon_2 - \left( 1 - e^{2 q d} \right) \epsilon_3 \right]} \nonumber \\
\epsilon_{bb} ( q ) &= \frac{e^{2 q d} \left( \epsilon_3 + \epsilon_2 \right) \left( \epsilon_1 + \epsilon_2 \right) - \left( \epsilon_3 - \epsilon_2 \right) \left( \epsilon_1 - \epsilon_2 \right)}{2 \left[ \left( 1 + e^{2 q d} \right) \epsilon_2 - \left( 1 - e^{2 q d} \right) \epsilon_1 \right]} \nonumber \\
\epsilon_{tb} ( q ) &= \frac{e^{2 q d} \left( \epsilon_3 + \epsilon_2 \right) \left( \epsilon_1 + \epsilon_2 \right) - \left( \epsilon_3 - \epsilon_2 \right) \left( \epsilon_1 - \epsilon_2 \right)}{4 e^{2 q d} \epsilon_2}
\end{align}
These expressions\cite{K11} interpolate the effective dielectric constant between $\lim_{q d \rightarrow 0} \epsilon_{tt} ( q ) = ( \epsilon_1 + \epsilon_3 ) / 2$ to $\lim_{q d \rightarrow \infty} \epsilon_{tt} ( q ) = ( \epsilon_1 + \epsilon_2 ) / 2$. The intra valley static polarizations of single layer and bilayer graphene are known\cite{WSSG06,SHS10,G11}. For bilayer graphene we use the expression which includes contribution from the four bands\cite{G11}.

The electronic excitations of the system, and their coupling to the electrons in each layer, can be described by ${\rm Im} v_{tt} ( \omega, q )$ and ${\rm Im} v_{bb} ( \omega, q )$. These quantities are shown in Fig.~\ref{excitations} and Fig.~\ref{excitations_2} for different values of the screening by the dielectric layers. The calculations include a finite broadening, in order to display the plasmon resonance. The charged plasmon has a large fraction of the spectral strength, and its dispersion depends on the screening constant. The acoustic plasmon has a dispersion
\begin{align}
\hbar \omega_{p acc} ( \vec{\bf q} ) &\approx \sqrt{\frac{4 e^2 \hbar  v_F k_F d}{\epsilon_1 + \epsilon_3}} \left| \vec{\bf q} \right|
\end{align}
For the parameters used in Fig.~\ref{excitations} and Fig.~\ref{excitations_2} the plasmon velocity is very close to the Fermi velocity, and its spectral weight in the effective interaction is small.

\begin{widetext}

\begin{figure}
\begin{center}
\includegraphics[width=0.4\columnwidth]{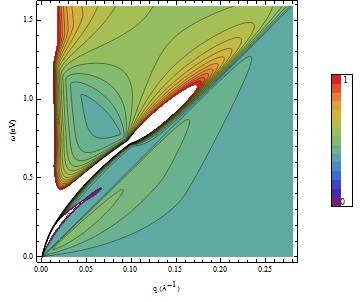}
\includegraphics[width=0.4\columnwidth]{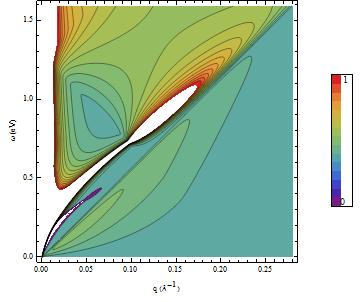}
\caption[fig]{\label{excitations} Imaginary part of the effective potential for an heteroestructure with $\rho_t = 4 \times 10^{13} {\rm cm}^{-2}, \rho_b = 10^{13} {\rm cm}^{-2}, \epsilon_1 = \epsilon_2 = \epsilon_3 = 4$, and $d = 10$ nm. Left: top layer. Right: bottom layer.}
\end{center}
\end{figure}

\begin{figure}
\begin{center}
\includegraphics[width=0.4\columnwidth]{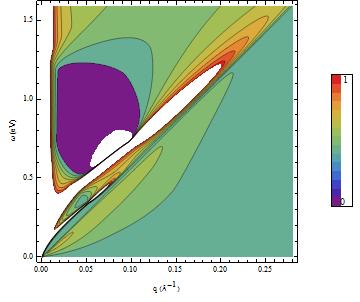}
\includegraphics[width=0.4\columnwidth]{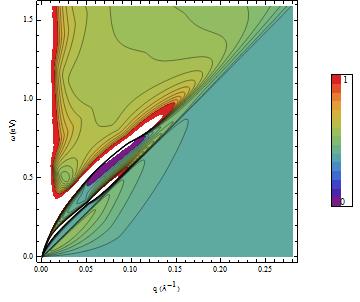}
\caption[fig]{\label{excitations_2} As in Fig.~\ref{excitations} but with $\epsilon_t = 10$. Left: top layer. Right: bottom layer.}
\end{center}
\end{figure}
\end{widetext}

\section{Kohn-Luttinger superconductivity in graphene heterostructures}
\subsection{Pairing interaction}

Superconductivity due to electron-electron interactions is proportional to the Fermi energy, $\epsilon_F$, times a factor which depends exponentially on the inverse of the coupling, $\lambda$, $T_c \approx 1.1 \epsilon_F e^{-1/\lambda_{e-e}}$. In materials where superconductivity is induced by phonons, the Fermi energy is replaced by the Debye frequency, $T_c \approx 1.1 \hbar \theta_D e^{-1/\lambda_{e-ph}}$. Because of the difference between the Fermi and Debye energies in a given material, electron-electron interactions have been considered a route to high critical temperatures, provided that the couplings are similar. The effective electron-electron interaction in an isolated 2DEG, however, is only weakly modulated as function of momentum $\vec{\bf q}$ in the relevant range $0 \le \left| \vec{\bf q} \right| \le 2 k_F$, where $k_F$ is the Fermi momentum\cite{C93}, and the Kohn-Luttinger instability, although present, leads to low critical temperatures.

In a two dimensional metallic system with an isotropic Fermi surface, the coupling constant $\lambda_n$ for a superconducting phase with an order parameter $\Delta_n ( \theta ) \propto \cos ( n \theta )$, is given by
\begin{align}
\lambda_n &= \frac{2 \rho ( \epsilon_F )}{\pi} \int_0^\pi V \left[ 2 k_F \sin \left( \frac{\theta}{2} \right) \right] \cos ( n \theta ) d \theta  \label{lambda}
\end{align}
where $\rho ( \epsilon_F )$ is the density of states per spin at the Fermi level, and $V ( \vec{\bf q} )$ is the effective electron-electron interaction. For superconductivity to occur, $\lambda_n$ must be negative. The critical temperature is given, approximately, by
\begin{align}
T_c &\approx 1.1 \epsilon_F e^{- \frac{1}{\lambda_n}}
\label{tc}
\end{align}

The calculation of $\lambda_n$ is simplified if the function $V ( q )$ has singularities or a non analytical dependence\cite{KL65} on $q = | \vec{\bf q} |$. For instance, when
\begin{align}
\lim_{| \vec{\bf q} | \rightarrow 0} V ( \vec{\bf q} ) = V ( 0 ) + V' ( 0 ) | \vec{\bf q} | + \cdots
\end{align}
we obtain
\begin{align}
\lim_{n \rightarrow \infty} \lambda_n = - \frac{2 \rho ( \epsilon_F ) V' ( 0 )}{\pi n^2} + O \left( \frac{1}{n^3} \right)
\end{align}
Hence, superconductivity at large angular momenta is possible when $V' ( 0 ) > 0$. In general, superconductivity is favored if $V ( | \vec{\bf q} | = 2 k_F ) > V ( 0 )$.

\begin{figure}
\begin{center}
\includegraphics[width=0.8\columnwidth]{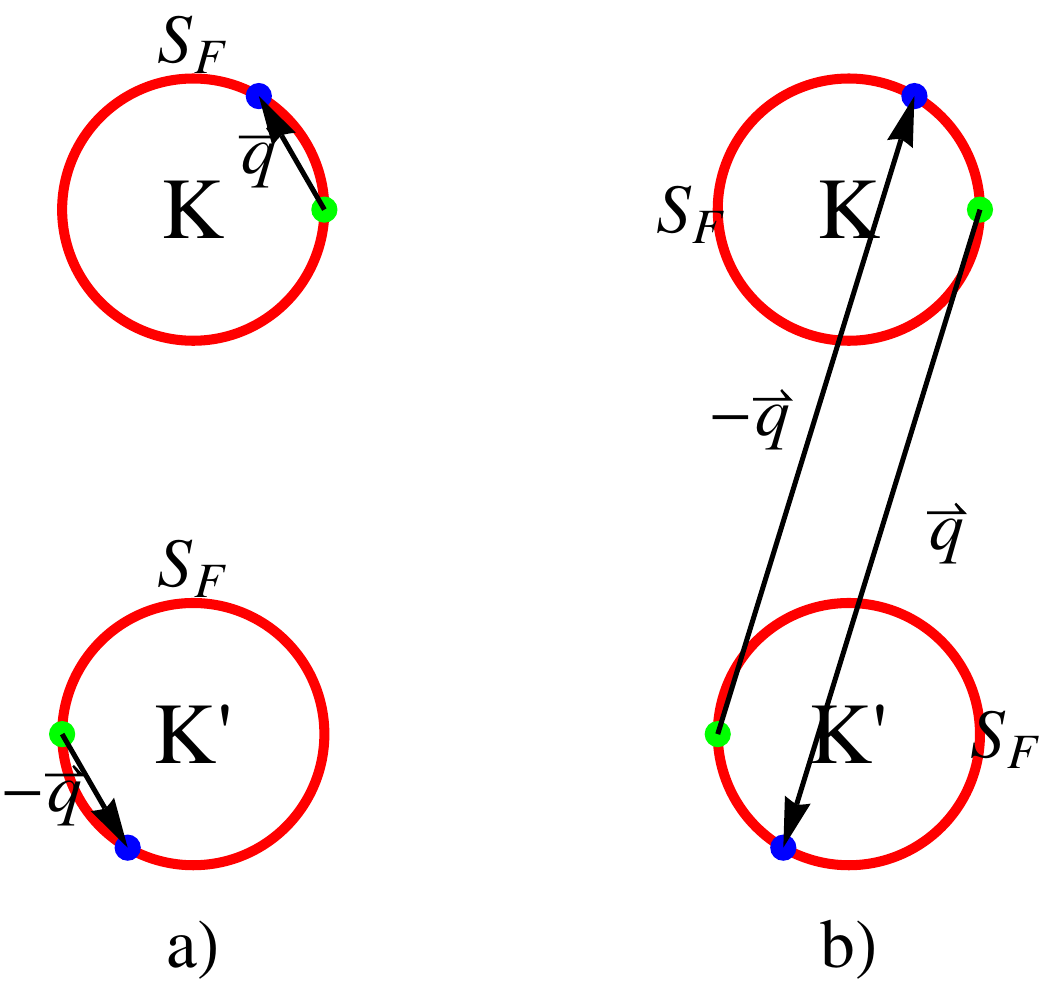}
\caption[fig]{\label{valleys}Sketch of the scattering processes of Cooper pairs considered in the text: a) Scattering within a given valley. b) Scattering between valleys.}
\end{center}
\end{figure}

\begin{figure}
\begin{center}
\includegraphics[width=0.8\columnwidth]{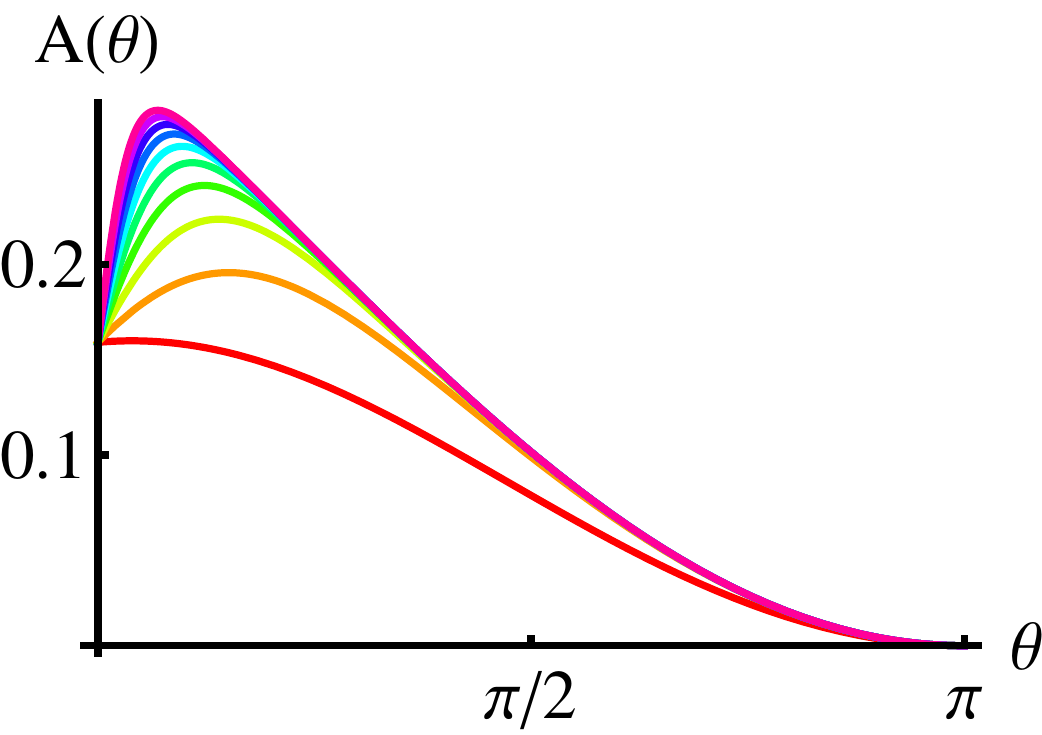}
\caption[fig]{ \label{atheta}Effective interaction, $A ( \theta ) = ( 2 / \pi )  \rho ( \epsilon_F^t ) V [ 2 k_F^t \sin ( \theta / 2 ) ]$ as function of $\theta$
 for an heterostructure made of two single layer graphene sheets with densities $\rho_t = \rho_b = 10^{13} {\rm cm}^{-2}$ and $d = 10, 40, 70, \cdots 280$ \AA (from bottom to top).}
 \end{center}
\end{figure}

\begin{figure}
\begin{center}
\includegraphics[width=0.8\columnwidth]{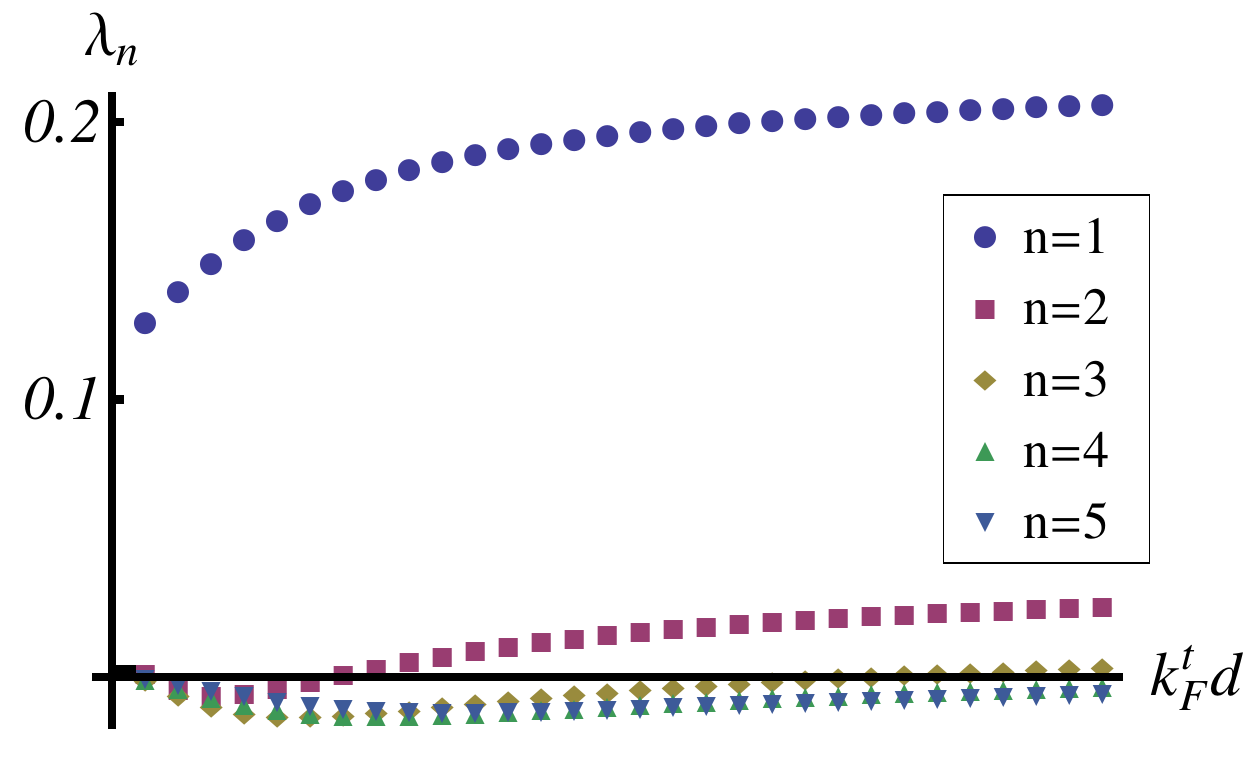}
\caption[fig]{\label{fig_lambda}Values of $\lambda_n$ as function of $k_F^t d$ for the same parameters as in Fig.~\ref{atheta}.}
 \end{center}
\end{figure}

\subsection{Pairing due to intravalley scattering}

 We assume that the distance between the chemical potential and the edge of the gap induced at the Dirac energy is larger than the gap itself, so that the pairing interaction does not couple the conduction and the valence bands. The densities of states per spin are
\begin{align}
\rho_{slg} ( \epsilon_F ) &= \frac{k_F}{\pi v_F} \nonumber \\
\rho_{blg} ( \epsilon_F ) &= \frac{1}{2 \pi \left( \frac{\gamma_1}{2} + \sqrt{\frac{\gamma_1^2}{4} + v_F^2 k_F^2} \right)}
\end{align}
The chiral nature of the wavefunctions in graphene changes the expression in eq.~\ref{lambda} into
\begin{align}
\lambda_n^{slg} &= \frac{2 \rho_{slg} ( \epsilon_F )}{\pi} \int_0^\pi V \left[ 2 k_F \sin \left( \frac{\theta}{2} \right) \right] \cos^2 \left( \frac{\theta}{2} \right) \cos ( n \theta ) d \theta  \nonumber \\
\lambda_n^{blg} &= \frac{2 \rho_{blg} ( \epsilon_F )}{\pi} \int_0^\pi V \left[ 2 k_F \sin \left( \frac{\theta}{2} \right) \right] \cos^2 \left( \theta \right) \cos ( n \theta ) d \theta
\label{lambda2}
\end{align}
A constant intervalley potential, $V ( \vec{\bf K} - \vec{\bf K}' )$ does not contribute to pairing in the situation considered here, where the gap is modulated within each valley. Higher order terms in an expansion in momenta around $\vec{\bf K} - \vec{\bf K}'$ fix the relative phase of the gaps in the two valleys.

As mentioned earlier, pairing is favored if the interaction potential is such that $V ( 2 k_F ) > V ( 0 )$. A finite value of $V ( 0 )$ is ensured by the metallic screening induced by the graphene layers. The value of $V ( 0 )$ is further reduced if the dielectric constant of one of the insulating layers is much larger than the other two, i. e. $\epsilon_3 \gg \epsilon_1 , \epsilon_2$, a situation which seems feasible in current experiments\cite{Petal09,CSM11}. Then, the screening in one layer changes from $( \epsilon_3 + \epsilon_1 )/2$ to $(\epsilon_2 + \epsilon_1 )/2$, as discussed before. For two graphene monolayers, the values of the dimensionless constants $\gamma_n$ are determined by $\alpha = e^2 / v_F , k_F^t d , k_F^b d , \epsilon_1 , \epsilon_2$ and $\epsilon_3$. We present in Fig.~\ref{atheta} calculations of the function $A ( \theta ) = 2 / \pi \rho_t ( \epsilon_F ) V [ 2 k_F^t \sin ( \theta / 2 ) ] \cos^2 ( \theta / 2 )$ for $\rho_t = \rho_b = 10^{13} {\rm cm}^{-2} , \epsilon_1 = \epsilon_2 = \epsilon_3  = 4,$ and different values of $d$. The value of $A ( \theta )$ raises for small $\theta$, with a slope which increases with $k_F^t d$. The orthogonality between wavefunctions of opposite momentum leads to $A ( \pi ) = 0$.

The coupling is enhanced using insulating layers with different dielectric constants. For $\epsilon_1 = \epsilon_2 = 1$ and $\epsilon_3 = 6$, and the densities and distances shown in Fig.~\ref{atheta} we find coupling constants of up to $\lambda_3 \sim - 0.055$. The critical temperatures are of order
\begin{align}
T_c &\approx \epsilon_F e^{-18}
\end{align}
These temperatures are too low to be observed experimentally.

\begin{figure}
\begin{center}
\includegraphics[width=0.8\columnwidth]{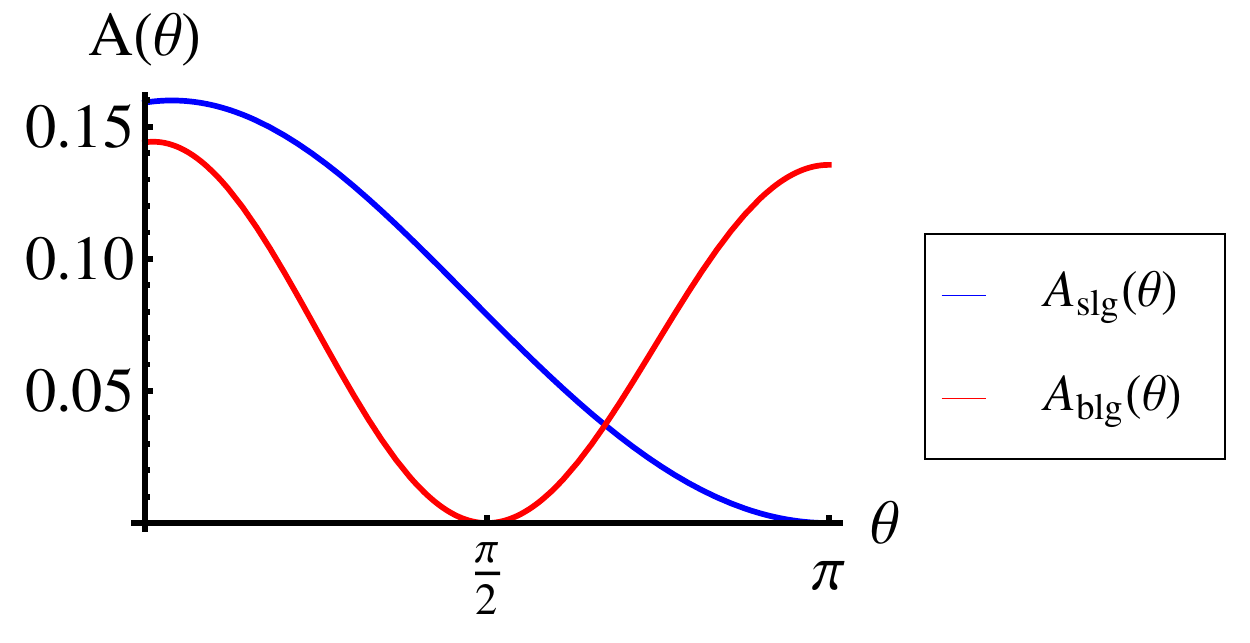}
\caption[fig]{\label{atheta_2}Effective interaction, $A ( \theta )$ for a graphene heterostructure made of single layer graphene and bilayer graphene with densities $\rho_{slg} = \rho_{blg} = 10^{13}$ cm$^{-2}$ and $d = 10$ \AA.}
 \end{center}

\end{figure}

These estimates do not change much for an heterostructure made of single layer and bilayer graphene. Fig.~\ref{atheta_2} shows results for a device with similar parameters as in Fig.~\ref{atheta}.

\subsection{Pairing due to intervalley scattering}
\begin{figure}
\begin{center}
\includegraphics[width=0.8\columnwidth]{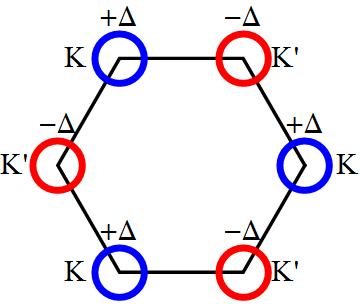}
\caption[fig]{\label{gap_BZ} Sketch of the gap with alternating signs in different valleys induced by intervalley scattering. The odd momentum pairing state, $\Delta(\vec{\bf k})=-\Delta(-\vec{\bf k})$, breaks time reversal and inversion symmetry.}
 \end{center}
\end{figure}

We consider now the scattering of Cooper pairs between valleys, see Fig.~\ref{valleys} b). We take $V ( \vec{\bf K} - \vec{\bf K'} ) \approx U \Omega$, where $U$ is the Hubbard repulsion between electrons on the same carbon atom, and $\Omega = 3 \sqrt{3} a^2 / 2$ is the area of the unit cell, and $a$ is the distance between carbon atoms. The next leading coupling, $V \sum_{ij n. n.} n_i n_j$, the repulsion between electrons in atoms which are nearest neighbors, does not contribute to the scattering process depicted in Fig.~\ref{valleys} b). Repulsive interactions between electrons in atoms at larger distances can be approximately included as an increase in the value of the Hubbard repulsion. The combination of weak intravalley and strong intervalley repulsion leads to a gap with alternating signs in different valleys, as sketched in Fig.~\ref{gap_BZ}. We assume that the gap has a constant value within each valley.

The value of the on site repulsion term in carbon $\pi$ orbitals, $U$, has been studied for a long time, by fitting model hamiltonians to the spectra of organic molecules\cite{PCR50,O64,BD69}. The value of $U$ is also a basic input in the analysis of excitonic peaks in carbon nanotubes\cite{PTA04,Jetal07}. Recent estimates\cite{VSCL10,Wetal11} suggest that a reasonable value of $U$ is $U \approx 10$ eV. This value is, approximately, $U \approx e^2/a$, or, alternatively, $U \Omega \approx 2 \pi e^2 / \vec{\bf K}$, where $\Omega = 3 \sqrt{3} a^2 / 2$ is the area of the unit cell, and $\left| \vec{\bf K} \right| = 4 \pi / \left( 3 \sqrt{3} a \right)$. The value $U \approx 10$ eV is also consistent with the value $U \approx 11$ eV used to fit the position of excitonic peaks in carbon nanotubes\cite{PTA04,Jetal07}.

We neglect the dependence of $U$ on momentum away from $\vec{\bf Q} = \vec{\bf K} - \vec{\bf K}'$. Then, pairing is only possible if $\Delta_K ( \theta )$ and $\Delta_{K'} ( \theta )$ are independent of $\theta$, and $\Delta_K = - \Delta_{K'}$. The coupling constant is
\begin{align}
\lambda_{iv} &=  \rho ( \epsilon_F ) \times \nonumber \\ &\times  \left\{ \frac{1}{\pi} \int_0^\pi V_{intra} \left[ 2 k_F \sin \left( \frac{\theta}{2} \right) \right] \cos^2 \left( \frac{\theta}{2} \right) d \theta - \frac{U \Omega}{2} \right\}
\label{lambdaiv}
\end{align}
Intervalley pairing is favored by the term
\begin{align}
\rho ( \epsilon_F ) \frac{U \Omega}{2} &= \frac{ 3 \sqrt{3} U k_F a^2}{4 \pi v_F} = \frac{\sqrt{3} U k_F a}{2 \pi \gamma_0}
\label{inter}
\end{align}
so that intervalley pairing is more likely at high densities.

We can obtain an order of magnitude estimate of the strength of the screened intravalley repulsion. In the limit $k_F d \rightarrow 0$, and assuming a single dielectric constant, $\epsilon_0$, we obtain
\begin{align}
\lim_{\theta \rightarrow 0} V \left[ 2 k_F \sin \left( \frac{\theta}{2} \right) \right] &\approx \lim_{q \rightarrow 0} \frac{2 \pi e^2}{\epsilon_0  + N_L N_f \frac{2 \pi e^2}{q} \frac{ k_F}{2 \pi v_F}} = \nonumber \\ &= \frac{2 \pi v_F}{N_L N_f k_F}
\label{intra}
\end{align}
where $N_L=2$ is the number of layers, and $N_f=4$ is the number of fermion species. When comparing the first and second term on the r. h. s. in eq.~\ref{lambdaiv} one needs to take into account that the absence of backward scattering, described by the factor $\cos^2 ( \theta / 2 )$ in the integrand of the intravalley contribution reduces by a factor of about $1/2$ that term. From eqs.~\ref{inter} and~\ref{intra} one can obtain an estimate of the range of parameters for which superconductivity is possible
\begin{align}
U \frac{3 \sqrt{3} a^2}{4} \gtrsim \frac{\pi v_F}{N_L N_f k_F}
\end{align}
Approximating $U \approx e^2 / a$, we obtain a minimum density above which superconductivity will appear
\begin{align}
\rho_{min} &= \frac{N_f k_F^2}{4 \pi} \gtrsim \frac{1}{a^2} \left[ \frac{4 \pi}{3 \sqrt{3} ( e^2 / v_F ) N_L N_f} \right]^2
\end{align}
For $\alpha = e^2 / v_F \approx 2.2$ and $a = 1.4$ \AA, we obtain $\rho_{min} \approx 8 \times 10^{13} {\rm cm}^{-2}$.  The value of the intervalley pairing term for $n =1.6 \times 10^{14} cm^{-2}$, is $\rho ( \epsilon_F ) U \Omega \approx 0.32$.

\begin{figure}
\begin{center}
\includegraphics[width=0.8\columnwidth]{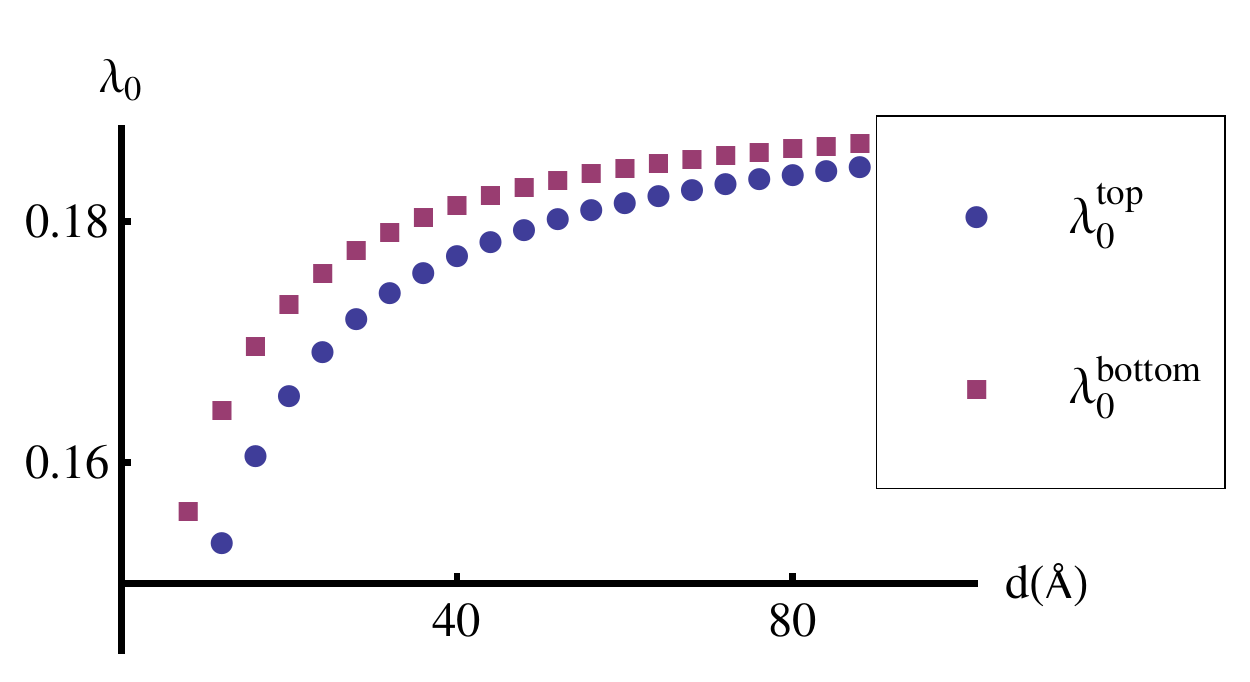}
\caption[fig]{\label{lambda_0_KL_1}Dependence of the parameter $\lambda_0$ in eq.~\ref{lambda0} on distance for an heterojunction with $\epsilon_1 = \epsilon_2 = \epsilon_3 = 4 , n_t = 10^{14} {\rm cm}^{-2}$ and $n_b = 2 \times 10^{14} {\rm cm}^{-2}$.}
 \end{center}
\end{figure}

A more precise analysis of the effects of intravalley interactions is described by the parameter
\begin{align}
\lambda_0 &=  \rho ( \epsilon_F ) \frac{2}{\pi} \int_0^\pi V_{intra} \left[ 2 k_F \sin \left( \frac{\theta}{2} \right) \right] d \theta
\label{lambda0}
\end{align}
The dependence of $\lambda_0^t$ and $\lambda_0^b$ on distance for a heterostructure made from two graphene single layers, and $\epsilon_1 = \epsilon_2 = \epsilon_3 = 4 , n_t = 10^{14} {\rm cm}^{-2}$ and $n_b = 2 \times 10^{14} {\rm cm}^{-2}$ is shown in Fig.~\ref{lambda_0_KL_1}. The critical temperatures of the two layers are $T_c^t \approx 0.6 {\rm K} , T_c^b \approx 100 {\rm K}$.

\begin{figure}
\begin{center}
\includegraphics[width=0.8\columnwidth]{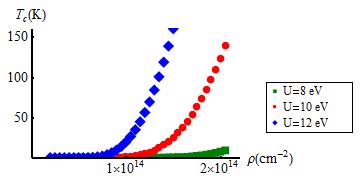}
\caption[fig]{\label{fig_tc}Dependence of the critical temperature on carrier density for an heterostructure made of two single layers with the same concentration, separated by a distance $d = 10$ \AA, and $\epsilon_1 = \epsilon_2 = \epsilon_3 = 4$.}
 \end{center}
\end{figure}

\begin{figure}
\begin{center}
\includegraphics[width=0.8\columnwidth]{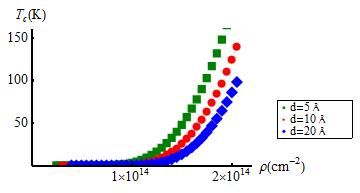}
\includegraphics[width=0.8\columnwidth]{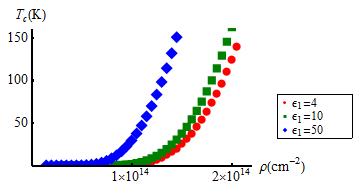}
\caption[fig]{\label{fig_tc_2} Dependence of $T_c$ on interlayer distance and dielectric constant. Top: Dependence on distance, for $U = 10$ eV and $\epsilon_1 = \epsilon_2 = \epsilon_3 = 4$. Bottom: Dependence on $\epsilon_1$, for $U = 10$ eV, and $d = 10$ \AA .}
 \end{center}
\end{figure}

The value of $T_c$ is  highly dependent on the input parameters, and is enhanced as the interlayer distance is reduced, or else in the presence of a substrate with a high dielectric constant, such as HfO$_2$, see Figs.~\ref{fig_tc} and~\ref{fig_tc_2}.

The mean field analysis reported here does not include phase fluctuations of the order parameter. In a two dimensional superconductor these fluctuations lower the critical temperature, and turn the phase transition into a Kosterlitz Thouless transition.  The change in critical temperature can be accounted for as a correction by numerical factor of order unity of the mean field value\cite{LT09}. The microscopic parameters which define the model have a weak influence on this factor.

\subsection{Electron-phonon coupling}
The leading contribution to the electron-phonon coupling in graphene is due to optical phonons at the $\Gamma$ and $K$ points of the Brillouin zone. These phonons can induce intra- and intervalley scattering, and optical phonons modify strongly the interatomic distances, and the nearest neighbor hopping elements. The electron phonon coupling induces an attractive interaction at all wavevectors, and favors the existence of a superconducting gap with the same sign in the two valleys\cite{ LS10,EE11}. Expanding around the Dirac points, the coupling can be parametrized by effective intra- and intervalley potentials, $g_{ph}^0 , g_{ph}^1$, which can be estimated from LDA calculations\cite{CM07}, or extracted from experiments\cite{YZKP07,BA08}. A tight binding estimate of this coupling\cite{G81,B08,BA08} is
\begin{align}
g_{ph}^{0,1} &\approx \frac{2 \beta^2 \hbar^2 v_F^2}{M_C \omega_{0,1}^2 a^4}  \times \Omega
\label{geph}
\end{align}
where $M_C$ is the mass of the carbon atom, $\omega_{0,1}$ is the frequency of the phonon, and $\beta = \partial \log ( \gamma_0 ) / \partial \log ( a ) \approx 2 - 3$. This estimate gives a dimensionless coupling, for $\beta = 2$, $\hbar \omega_0 =0.2$ eV, which is doubly degenerate, and $\hbar \omega_1 = 0.17$ eV,
\begin{align}
\lambda &= \frac{2 g_{ph}^0 + g_{ph}^1}{2} \times \frac{k_F}{\pi v_F} \approx 5.05 \sqrt{n} \times 10^{-9} {\rm cm}
\end{align}
The value of this coupling is in reasonable agreement with LDA calculations\cite{BA08}, where the numerical factor is 5.5.

\begin{figure}
\begin{center}
\includegraphics[width=0.8\columnwidth]{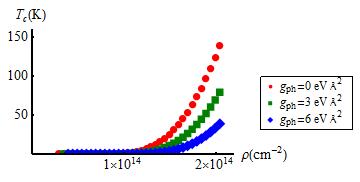}
\caption[fig]{\label{eph} Dependence of $T_c$ on  the electron phonon coupling, $g_{ph}$. The remaining parameters are $U = 10$ eV, d = $10$ \AA , $\epsilon_1 = \epsilon_2 = \epsilon_3 = 4$. }
 \end{center}
\end{figure}

The total coupling obtained from eq.~\ref{geph} is $g_{ph} = 2 g_{ph}^0 + g_{ph}^1 \approx 2.3$ eV \AA$^2$. This coupling reduces the effect of the Coulomb interaction. The resulting effect on the value of $T_c$ is shown in Fig.~\ref{eph}.

\subsection{Andreev states and pair breaking effects in the superconducting phase}
The superconducting phase studied in the last subsection shows a gap which is constant within each valley and has opposite signs in the two valleys. Elastic scattering within each valley does not change this phase, while intervalley scattering is pair breaking. Defects which induce intervalley scattering induce Andreev states inside the superconducting gap. As an example, we consider a local perturbation, described by a shift of the onsite energy of a single $\pi$ orbital, which we denote $\delta \epsilon$. The local density of states in the presence of the perturbation is
\begin{align}
\hat{G} ( \omega ) &= \left[ \hat{G}_0^{-1} ( \omega ) - \left( \begin{array}{cc} \delta \epsilon &0 \\ 0 &- \delta \epsilon \end{array} \right) \right]^{-1}
\label{green}
\end{align}
where
\begin{align}
\hat{G}_0 ( \omega ) &= \hat{G}_0^K ( \omega ) + \hat{G}_0^{K'} ( \omega )
\end{align}
\and
\begin{align}
\hat{G}_0^K &\approx i \rho ( \epsilon_F ) \frac{\omega}{\sqrt{\omega^2 - \Delta_K^2}} \times \nonumber \\ &\times \left( \begin{array}{cc} \frac{1}{2} \left( 1 + \frac{\sqrt{\omega^2 - \Delta_K^2}}{\omega} \right) &\frac{\Delta_K}{\omega} \\ \frac{\Delta_K}{\omega} &\frac{1}{2} \left( 1 - \frac{\sqrt{\omega^2 - \Delta_K^2}}{\omega} \right) \end{array} \right)
\end{align}
where $\rho ( \epsilon_F ) = 2 k_F / ( \pi v_F ) \times \Omega$ is the density of states at the Fermi energy and $\Omega$ is the area of the unit cell. Using $\Delta_K = - \Delta_{K'}$, we find that the Green's function in eq.~\ref{green} has poles at
\begin{align}
\omega &= \pm \frac{\Delta}{1 + \rho ( \epsilon_F ) \delta \epsilon}
\end{align}
A vacancy can be described as the limit $\delta \epsilon \rightarrow \infty$. In this limit, we find two midgap states, at $\omega = 0$. Impurities which induce intervalley scattering induce pair breaking and reduce the gap. Superconductivity will be completely suppressed for $\Delta \approx v_F / \ell_{K-K'}$, where $\ell_{K-K'}$ is the elastic mean free path associated to intervalley scattering.


\begin{figure}
\begin{center}
\includegraphics[width=0.6\columnwidth]{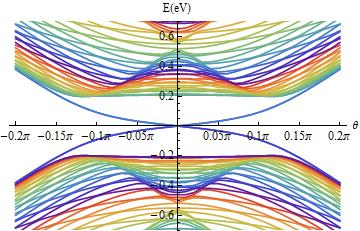}
\includegraphics[width=0.6\columnwidth]{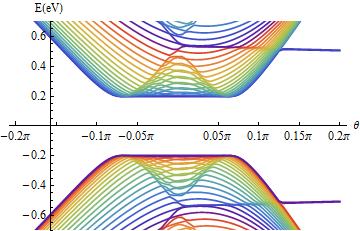}
\caption[fig]{\label{bands} Subbands of a graphene ribbon with a superconducting gap which has opposite signs in the two valleys. In order to illustrate better the main features, the Fermi energy is $\epsilon_F = 0.5$ eV, and the gap is $\Delta = 0.2$ eV. Top: armchair edge. Bottom: zigzag edge.}
 \end{center}
\end{figure}

In a similar way as in other multiple band materials, Andreev states can be induced at surfaces\cite{BG01}. Alternatively, surface states can be seen as arising from topological features of the bulk band. In order to describe effects due to the breaking of translational symmetry, it is convenient to write the Bogoliubov-de Gennes hamiltonian in a real space basis. A superconducting gap which changes sign between valleys in the honeycomb lattice can be described by a tight binding model with purely imaginary hopping terms which mix electron and hole states at sites the same sublattice, following the same method used in the formulation of the Haldane model\cite{H88}. The tight binding model contains four entries per lattice site. In the absence of magnetic impurities, the model can be split into two equivalent systems. In the presence of electron-hole symmetry, with the chemical potential at the Dirac point, each of these models can be further split into two Haldane models with opposite Chern numbers. Thus, the odd pairing considered here, in the presence of full electron-hole symmetry can be seen as the sum of independent topological superconductors. For finite chemical potentials, however, this reduction cannot be done, and the superconducting phase has no topologically protected Andreev states at any edge.

In the absence of general topological arguments, Andreev states can be expected only at surfaces where the valleys are hybridized. Fig.~\ref{bands} show the subbands of superconducting nanoribbons with different orientations. In agreement with the previous analysis, subgap states exist for armchair nanoribbons, while they are absent in zigzag nanoribbons. Note that the edge states at the zigzag edge are at the Dirac energy, shifted away from the Fermi level and the superconducting gap.

Another type of defects which lead to the hybridization of the two valleys are topological defects, like pentagons or heptagons. These defects can be described at long wavelengths as vortices acting on two effective Dirac equations with a flux $\Phi = \pm (e h) / ( 2 c )$, where the two opposite signs ensure time reversal invariance\cite{GGV92,GGV93}. A loop around a defect exchanges the two valleys. A superconducting gap with opposite signs in the two valleys changes sign around such a loop. The resulting system can be viewed as two "one fourth" vortices, with each of them having a midgap state\cite{RG00,I01}, which can be split by residual interactions\cite{H10}, see also\cite{GCSV12}. In the presence of a magnetic field, topological defects can pin vortices with $\Phi = ( e h ) / ( 2 c )$. This configuration leads to an effective "half vortex" in one of the two Dirac equations, and to a single midgap state, which can be described as a Majorana particle.

\subsection{Kohn-Luttinger superconductivity and excitonic condensation}
The analysis discussed here bears a number of similarities to the study of excitonic condensation in an artificial graphene bilayer when the number of electrons in one layer equals the number of electrons in the other layer\cite{LS08,MBSM08,KE08}. If one assumes isotropic Fermi surfaces in both layers, the pairing strength can be described by a parameter similar to $\lambda_{iv}$ in eq.~\ref{lambdaiv}, except that the first term is negative, and the second term, associated to short range interactions, is absent.

Long range repulsive interactions, which are detrimental for the Kohn-Luttinger mechanism of superconductivity, are the source of excitonic pairing in an artificial bilayer with electrons in one layer and holes in the other. An estimate of the effect of long range interactions is given in eq.~\ref{intra}, and a similar value was obtained in\cite{KE08} in relation to the excitonic transition.

An important difference between the Kohn-Luttinger mechanism of superconductivity and the excitonic transition is the role of a misalignment between the two layers and the trigonal distortion of the Fermi surface at high carrier concentrations. Superconductivity is due to the existence of Cooper pairs where the two carriers involved are related by time reversal symmetry, and they are not affected by deformations and misalignments of the Fermi surfaces (see, however\cite{symmetry}). The electron and the hole which are paired and give rise to the excitonic condensate arise from the same valley, in order for the exciton to have zero total momentum. A rotation of the Fermi surface, combined with trigonal warping, plays the same role as a pair breaking perturbation in a superconducting phase. We will not consider here the enhancement of the superconducting transition temperature due to fluctuations of the order parameter, a possibility which has been studied in exciton condensation\cite{LOS12,SPM12}.

\section{Discussion}
The modulation of purely repulsive electron-electron interactions leads to superconductivity with an anisotropic order parameter\cite{KL65}. Pairing by this mechanism in isotropic metals is favored when the interaction shows a strong momentum dependence, $V ( 2 k_F ) \gg V ( 0 )$ and/or singularities at finite momenta. Pairing is be enhanced in anisotropic systems, if the interaction which couples regions of the Fermi surface with a high density of states is larger than the interaction at small momenta. Superconductivity induced by repulsive interactions which couple different valleys has been proposed for the iron pnictides\cite{MSJD08,CEE08}, and it is related to the "Van Hove scenario" extensively studied in relation to the cuprate superconductors\cite{GGV96}.

Graphene heterostructures allow for the modification of the long wavelength part of the electron-electron interaction. The absolute value of the interaction can be reduced, making it smaller than the short range repulsion, and the dependence on momentum $\vec{\bf q}$ can be modulated. The lowering of the average value of the long wavelength interactions, combined with a significant local repulsion, can lead to superconductivity at experimentally accessible temperatures, for carrier densities above $5 \times 10^{13} {\rm cm}^{-2}$, and interlayer distances of a few nanometers. The embedding of the device in a dielectric medium with a large dielectric constant increases the tendency towards pairing.

Superconductivity for an odd momentum pairing state, where $\Delta_{\vec{\bf k}}= - \Delta_{- \vec{\bf k}}$, is  unlikely  to occur in a single layer near a metallic surface\cite{ON07,Vetal08b,Petal09b,Getal11}. The metal can shift the chemical potentia of graphene by 0.1-0.4 eV and a charge transfer of $10^{12}-10^{13} \rm{cm}^{-2}$ is possible.  Nevertheless, the hybridization with the electrons of the metal will produce backscattering, which  breaks Cooper pairs and should completely suppress superconductivity when the broadening of the graphene states  $\Gamma$ is of order $\Gamma \gtrsim k_B T_c$.

In the graphene heterostructures considered here, superconductivity with a constant gap in each valley, and opposite sign in different valleys, is possible. This symmetry implies the existence of rise of triplet Cooper pairs with $s_z = 0$. The superconducting phase is gapped, and short range scalar disorder is be pair breaking. Surfaces and topological defects can induce midgap states.

We have also found a second superconducting instability, induced solely by the intravalley interaction.  This instability resembles closely the original superconducting instability of an isotropic electron gas\cite{KL65}. As in that case, it occurs at temperatures too low to be accessible experimentally.

Accurate predictions of the value of the critical temperature of a superconductor are extremely difficult, as the value depends exponentially on parameters which are not known with great precision. Our estimates depend on the electrostatic repulsion between carriers in different layers of the heterostructure, and on the value of the short range interaction.
The graphene heterostructures studied so far show significant effects due to long range interactions between layers. The analysis of Coulomb drag at high temperatures and high carrier concentrations  seems consistent with models based on the Random Phase Approximation, which is also used here.

In order for superconductivity to occur, the short range repulsion must be larger than the long wavelength interaction. The value of $U$, the onsite interaction in graphene, is difficult to determine experimentally. Calculations for graphene, aromatic molecules, and conjugated polymers, give values which cluster consistently in the range $U \sim 8 - 12$ eV, although larger and smaller values have been proposed. For $U \gtrsim 9$ eV, superconductivity at accessible temperatures can be expected for carrier concentrations $\rho \gtrsim 5 \times 10^{13} {\rm cm}^{-2}$. Longer range interactions, still comparable with the lattice constant, will probably enhance the effect of intervalley scattering, increasing the value of $T_c$.

Pairing is very sensitive to the dielectric constant of the insulating layers. A lesser source of uncertainty is the use of the Fermi energy as the energy scale over which pairing takes place. This energy enters in the prefactor of the expression for $T_c$. A change in this parameter does not alter the balance between long range and short range interactions which is responsible for the existence of superconductivity. Fluctuations of the superconducting phase depress the critical temperature, but their effect is not as critical as the details of the electron-electron interaction.

\section{Acknowledgements}
The authors acknowledge the hospitality of the KITP, Santa Barbara. This work is partially
supported by the National Science Foundation under Grant No. NSF PHY11-25915. This work has been funded by the Spanish MICINN (FIS2008-00124, CONSOLIDER CSD2007-00010), and ERC, grant 290846. The authors appreciate a number of useful conversations with J. Gonzalez and P. San-Jose.

\bibliography{artificial_bilayer}
\end{document}